\documentclass[aps,prb,twocolumn,superscriptaddress,showpacs]{revtex4}
\usepackage{graphics}
\usepackage{subfigure}
\usepackage{longtable}
\def\LOFA{LaOFeAs}
\def\LOFFA{{LaO$_{1-x}$F$_x$FeAs}}
\def\LOFFP{{LaO$_{1-x}$F$_x$FeP}}

\begin{document}

\preprint{}

\title{Proximity of antiferromagnetism and superconductivity in \LOFFA: effective Hamiltonian from ab initio studies}

\author{Chao Cao}
\affiliation{Department of Physics, University of Florida, Gainesville, FL 32611, U.S.A.}
\affiliation{Quantum Theory Project, University of Florida, Gainesville, FL 32611, U.S.A.}
\author{P. J. Hirschfeld}
\affiliation{Department of Physics, University of Florida, Gainesville, FL 32611, U.S.A.}
\author{Hai-Ping Cheng}
\email[Corresponding author, e-mail: ]{cheng@qtp.ufl.edu}
\affiliation{Department of Physics, University of Florida, Gainesville, FL 32611, U.S.A.}
\affiliation{Quantum Theory Project, University of Florida, Gainesville, FL 32611, U.S.A.}

\date{\today}

\begin{abstract}
We report density functional theory calculations for the parent
compound \LOFA\ of the newly discovered 26K Fe-based
superconductor \LOFFA.  We find that the ground state is an
ordered antiferromagnet, with staggered moment about 2.3$\mu_B$,
on the border with the Mott insulating state. 
We fit the bands crossing the Fermi surface, derived from Fe and
As,  to a tight-binding Hamiltonian using maximally localized
Wannier functions on Fe 3d and As 4p orbitals. The model
Hamiltonian accurately describes the Fermi surface obtained via
first-principles calculations. Due to the evident proximity of
superconductivity to antiferromagnetism
and the Mott transition, we suggest that the system
may be an analog of the electron doped cuprates, where antiferromagnetism
and superconductivity coexist.

\end{abstract}

\pacs{74.70.-b,74.25.Ha,74.25.Jb,74.25.Kc}

\maketitle

The recent discovery of superconductivity with onset temperature
of 26K in \LOFFA\ \cite{LOFA_JACS} has generated considerable
interest because of a number of unusual aspects of this material.
First, with the exception of some of the A-15 materials, Fe is
never found in superconductors at zero pressure (although Fe
itself superconducts at 10GPa). Second,  both  ferromagnetic {\it
and} antiferromagnetic fluctuations are apparently present in the
material, suggesting possible analogies to ternary rare earth,
heavy fermion, borocarbide, ruthenate and cuprate superconductors.
Finally, the discovery of superconductivity at the relatively high
critical temperature of 26K implies that a new pairing mechanism
may be in play.  The analog system \LOFFP\, has a critical
temperature of 7K, so there appears to be a new class of
superconducting materials with no obvious limit on $T_c$.

Naively, ferromagnetic order is inimical to superconductivity since the
exchange field of the ferromagnetic ion breaks singlet pairs.
Ferromagnetic order was therefore  found only very recently to
coexist with superconductivity in UGe$_2$\cite{Saxena} and
URhGe\cite{Aoki}, in situations where the order is quite weak.
Coexistence of antiferromagnetic order, on the other hand, is less
pairbreaking if the coherence length is much larger than the
wavelength of the magnetic modulation, which is generally the
case.  Thus many examples of superconductors coexisting with
antiferromagnetic order are known, and have been recently
reviewed\cite{EPugh}.

The structure of the new material consists of LaO layers
sandwiching a layer of FeAs, and doping with F appears to occur on
the O sites.  Since the Fe is arranged in a simple square lattice,
the analogy with the cuprates, where electrons hop on a square
lattice and doping occurs via a nearby oxide charge reservoir
layer, is tempting to draw.  Early electronic structure
calculations for both the P\cite{LOFP_PRB} and the new As
materials\cite{LOFA_Singh,LOFA_CAS,LOFA_Haule} have presented a somewhat
different picture, however.  While experimentally a low charge
density was measured for this material\cite{LOFA_CAS_HALL}, these
calculations suggest that five bands, of primarily Fe-As
character, cross the Fermi level and give rise to a multi-sheeted,
quasi-2D material. No evidence for long-range order was found in
these studies, although proximity to both antiferromagnetic and
ferromagnetic ordered states was noted.

The weak coupling of the LaO layers to the FeAs layers found here
and in previous works also suggests that insight may be gained by
examining iron monoarsenide FeAs, a layered metallic
helimagnet\cite{FA_Selte} with large pitch angle in the zincblende
structure with low-temperature effective moment $\sim0.5\mu_B$.
The electronic structure and spin moment for this compound have
been calculated by density functional theory\cite{FeAs_DFT}. While
the theory is successful in the sense that an antiferromagnetic
ground state is found, it does not distinguish the complicated
magnetic structure and finds a moment of order $\sim 2\mu_B$.

To understand the electronic structure properties of the new
Fe-based superconductors and the interplay with structure and
magnetic states, we have performed first-principles density
functional theory simulations on the undoped and $x$=0.0625 as well 
as $x$=0.125 doped
\LOFFA. Most of the reported results on electronic structure and model
Hamiltonian have been obtained using the PWSCF package \cite{PWSCF},
which employs a plane-wave basis set and ultrasoft
pseudopotentials \cite{VANDERBILT_PP}. We have also used VASP \cite{VASP_1,VASP_2}
to confirm our calculations when the same calculations can be performed, as
described in the following sections in detail. The local spin
density approximation (LSDA) and generalized gradient
approximation (GGA) of Perdew, Burke and Ernzerhof (PBE)
\cite{PBE_xc} potentials have been incorporated for the
simulation. For density of states (DOS) calculations, we have used
a 16$\times$16$\times$8 Monkhorst dense grid \cite{Monkhorst_kgrid}
to sample the Brillouin zone; while for structural relaxation and self-consistent calculations, a
8$\times$8$\times$4 Monkhorst grid has been used. All structures
have been fully optimized until internal stress and forces on each
atom are negligible. Our GGA+U calculations have been performed via
the VASP code. The existing literature uses an on-site Columb energy
that varies from 4.0 eV to 6.9 eV \cite{LDAU_1,LDAU_2,LDAU_3}, and
we have explored the parameter space within the range $U$=2.0-5.0 eV and $J$=0.89 eV on Fe in our calculations.

\begin{figure}[htp]
  \centering
  \subfigure[]{
    \scalebox{0.40}{\rotatebox{270}{\includegraphics{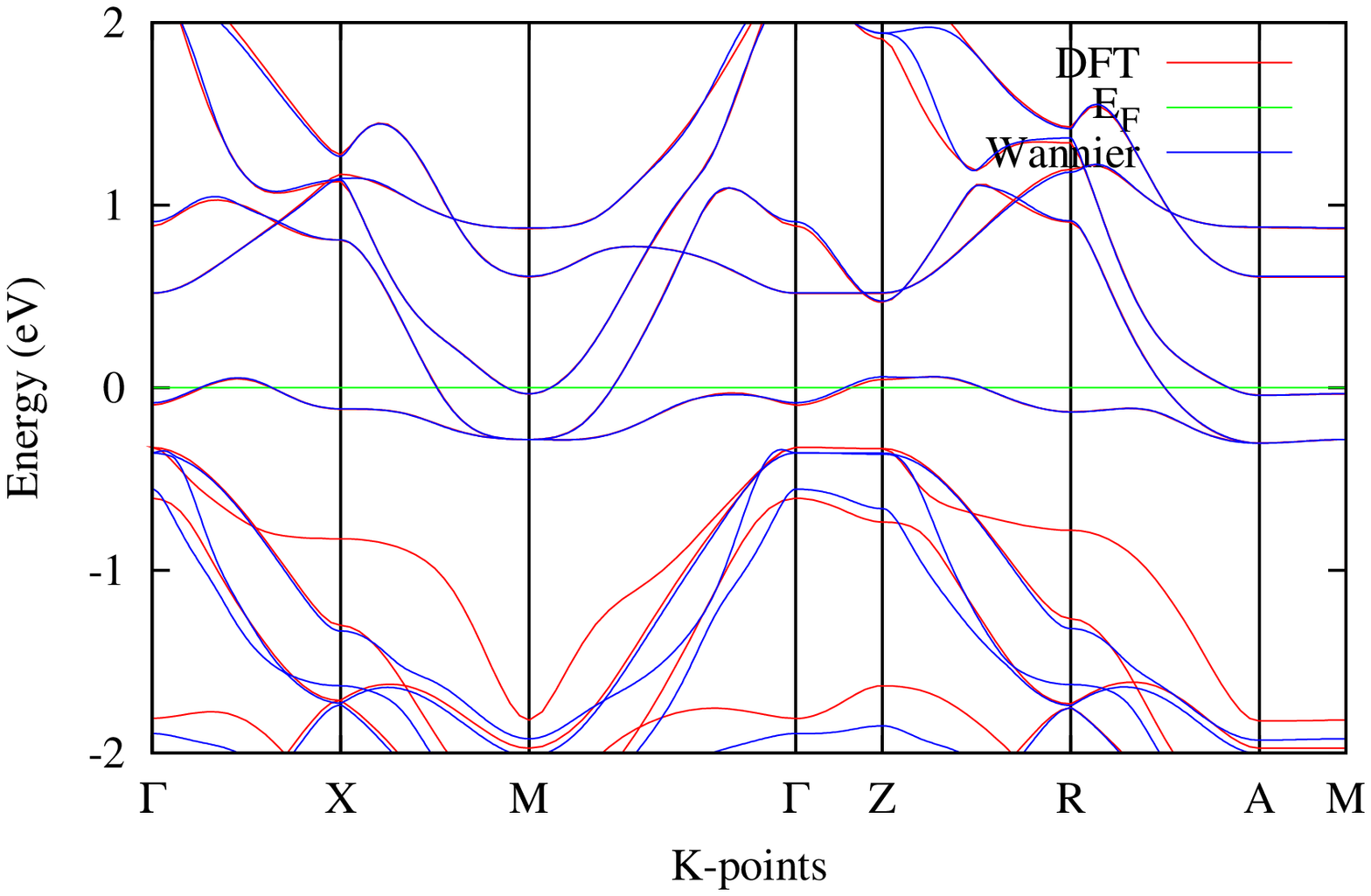}}}
    \label{fig_afm_band}
  }
  \subfigure[]{
    \scalebox{0.40}{\rotatebox{270}{\includegraphics{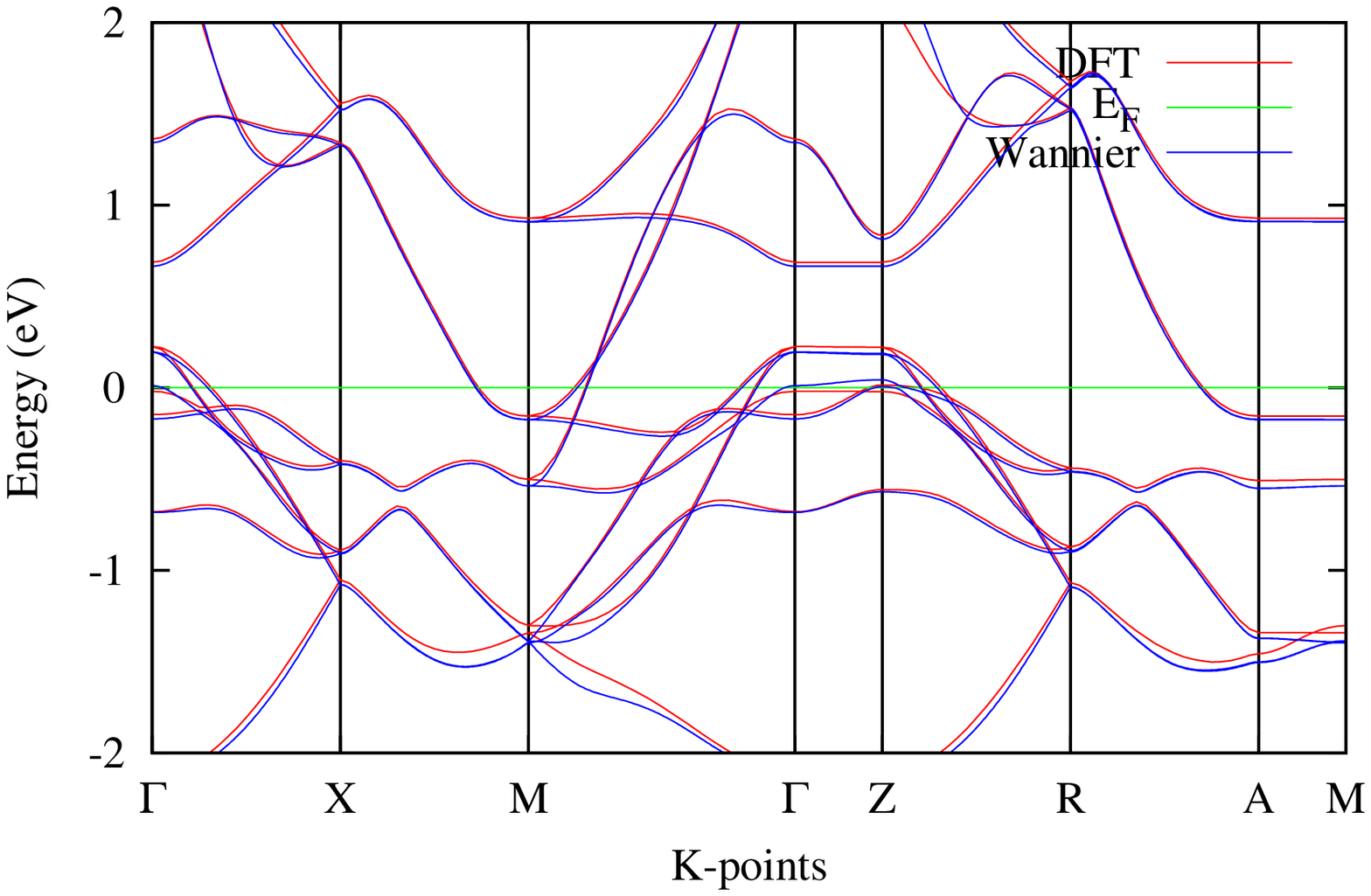}}}
    \label{fig_pm_band}
  }
  \caption{(Color online)Undoped LaOFeAs band structure of (a) AFM state and (b) PM state. Red
lines represent DFT calculation results; blue lines are band
structures reconstructed from the tight-binding model using maximally
localized Wannier functions (MLWF). Since spin-up and spin-down
bands are degenerate for AFM state, we plot only spin-up bands
here. For both figures, Fermi energies are indicated by the green line
at 0 eV.}
  \label {fig_band_structure}
\end{figure}

Our calculations show an unambiguous antiferromagnetic (AFM)
ground state with staggered moment 2.3 $\mu_B$ for undoped
LaOFeAs, which is 84 meV per Fe lower than
paramagnetic (PM) and ferromagnetic (FM) states. The energy
difference between the latter two is found to be negligible. In fact,
the FM state has a very small magnetic moment ($\sim$0.05 per Fe); 
therefore it can be regarded as a PM state. The
AFM ground state has been confirmed by independent VASP
calculations using the projector augmented wave (PAW) method \cite{PAW}. The
optimized structure has a lattice constant of $a$=4.0200 \AA\ and
$c$=8.7394 \AA; and the bond lengths for Fe-As and La-O are 2.35
\AA\ and 2.40 \AA\, respectively. For reference, the paramagnetic
state has an optimized lattice constant of $a$=3.9899 \AA\ and
$c$=8.6119 \AA, while the bond lengths for Fe-As and La-O are 2.34
\AA\ and 2.33 \AA, respectively. Both the AFM and PM band
structures are shown in Fig. \ref{fig_band_structure} with red
curves. In both states, a small dispersion  along the $c$-axis (from
$\Gamma$ to Z and from A to M) indicates interactions between
layers are weak, and thus the separation of the structure into LaO
and FeAs layers is possible. The PM state band structure
reproduces previous DFT calculation results
\cite{LOFA_Singh,LOFP_PRB}, exhibiting 5 bands across the Fermi
level. The AFM state band structure is qualitatively different,
exhibiting only 3 bands across $E_F$. 
In VASP calculations, AFM states are 14 meV per Fe lower than PM 
and FM states. Similar to PWSCF calculations, the band structures
of the AFM state are very different from the PM state. These results
indicate the delicacy of the magnetic states in this system, and
that the magnetism strongly affects the electronic structure. 

\begin{figure}[htp]
  \centering
  \scalebox{0.4}{\rotatebox{270}{\includegraphics{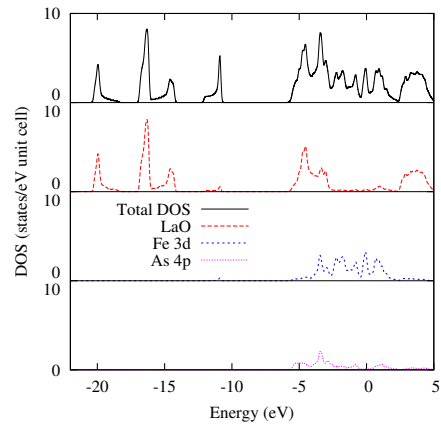}}}
  \caption{(Color online)DOS (top panel) and PDOS (FeAs and LaO planes: middle panel and bottom panel, respectively) of undoped LaOFeAs. Since the spin-up and spin-down
states are degenerate for AFM state, we plot spin-up states only. The Fermi
level is aligned to 0.0 eV.}
  \label{fig_pdos_afm}
\end{figure}

\begin{figure}[htp]
  \centering
    \scalebox{0.35}{\rotatebox{270}{\includegraphics{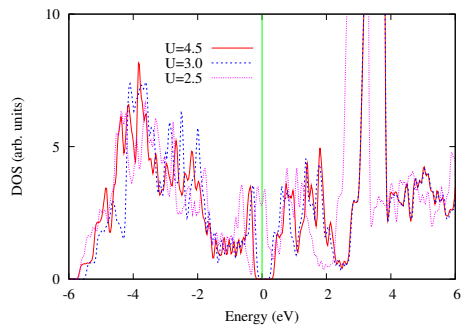}}}
  \caption{(Color online)DOS calculated from GGA+U with different $U$s, 2.0, 
3.0, and 4.5 eV. $E_F$ is aligned to 0.0 eV in all cases. Due to degeneracy,
only the $\alpha$-spin DOS is plotted.}
  \label{fig_ldau}
\end{figure}

To further examine and confirm these findings, we have performed
two series of additional calculations. First, we have performed
GGA+U calculations using VASP. Haule {\it et. al} used a dynamical mean
field theory (DMFT)-LDA approach, and found that a critical value
of $U$=4.5 eV led to a Mott transition with a gap at the Fermi
surface. We have calculated the electronic structure using VASP's
implementation of GGA+U, and also find a Mott transition for the
\LOFA\ system at a critical $U\sim$ 3 eV for Fe. Note that lower 
bound of the empirical value of $U$ chosen in calculation is 3.5-4.0 eV
for Fe d orbitals (\cite{LDAU_1}).
The ground state is found to be always AFM for all tested $U$ within 0.0-5.0 eV 
in our calculations, but the DOS changes dramatically (Fig. \ref{fig_ldau}). 
A Mott gap of about 1.0 eV is observed in the
GGA+U calculation at $U=4.5$ eV. Experimentally, it is observed
that below 100K, the resistivity of undoped LaOFeAs increases
when temperature decreases, but appears to remain
metallic\cite{LOFA_JACS}, suggesting that the system is in fact on
the edge of a Mott transition.\cite{LOFA_Haule}.

Second, we have investigated bulk FeAs. The AFM state is again
found to be the ground state, in agreement with experiment \cite{FA_Selte},
as well as with previous calculations based on full-potential 
linearized augmented planewave (FLAPW) method \cite{FeAs_DFT}. In
addition, an isolated layer of FeAs  has also been simulated, and
the system again has an AFM ground state. The calculated AFM
states of bulk and the isolated layer are 53 meV and 408 meV per
Fe atom lower than their PM states, respectively. Compared to FLAPW calculations, 
our results have shown a smaller energy difference between AFM and PM state, indicating
that our calculations do not have an artificial bias for the AFM state.

The electron DOS derived from the AFM band
structure and the corresponding projected DOS (PDOS) onto LaO and
FeAs planes are presented in Fig. \ref{fig_pdos_afm}. It is clear
from the PDOS analysis that Fe 3d orbitals dominate the DOS around
$E_F$ (-2 to 2 eV relative to $E_F$) and from -12 to -10 eV; whereas
DOS below -13 eV is almost completely derived from LaO layers.
This clear separation in the DOS confirms that this material can
be viewed as a layered structure. However, both LaO and FeAs
layers contribute approximately the same from -6 to -2 eV,
suggesting a hybridization between layers within this energy
window.  For AFM state, the calculations give a 2.30 $\mu_B$ local
magnetic moment on Fe by integrating the PDOS to $E_F$, which is
similar to the value obtained from calculations for the bulk FeAs
crystal.
\begin{figure}[htp]
  \centering
  \subfigure[]{
    \scalebox{0.35}{\includegraphics{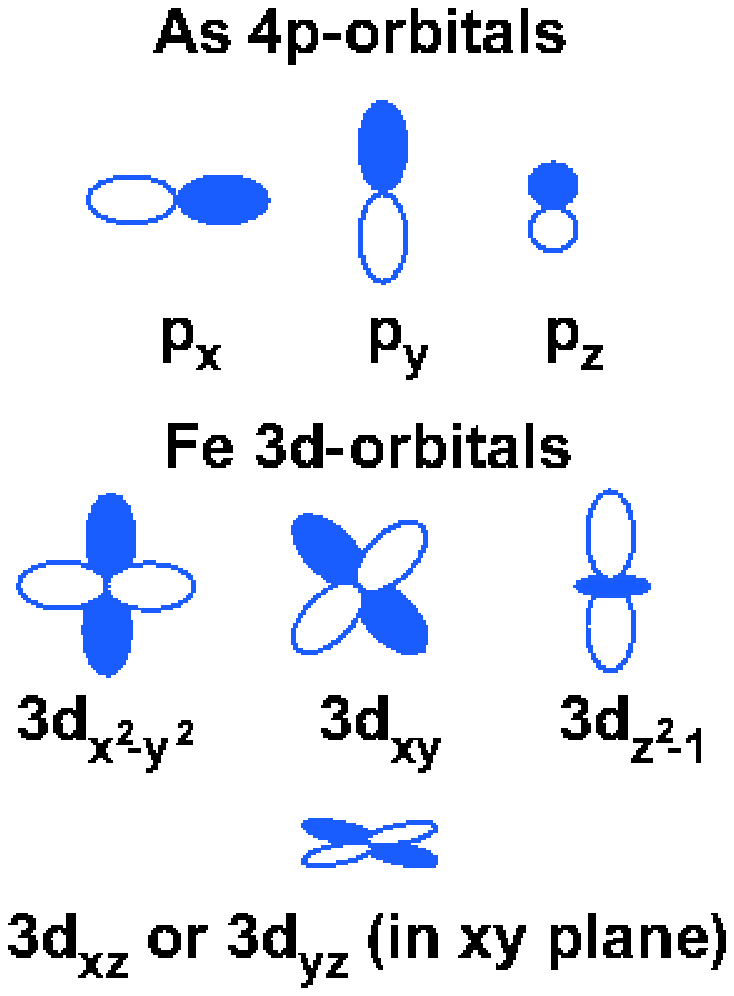}}
    \label{fig_orb_diagram}
  }
  \subfigure[]{
    \scalebox{0.20}{\includegraphics{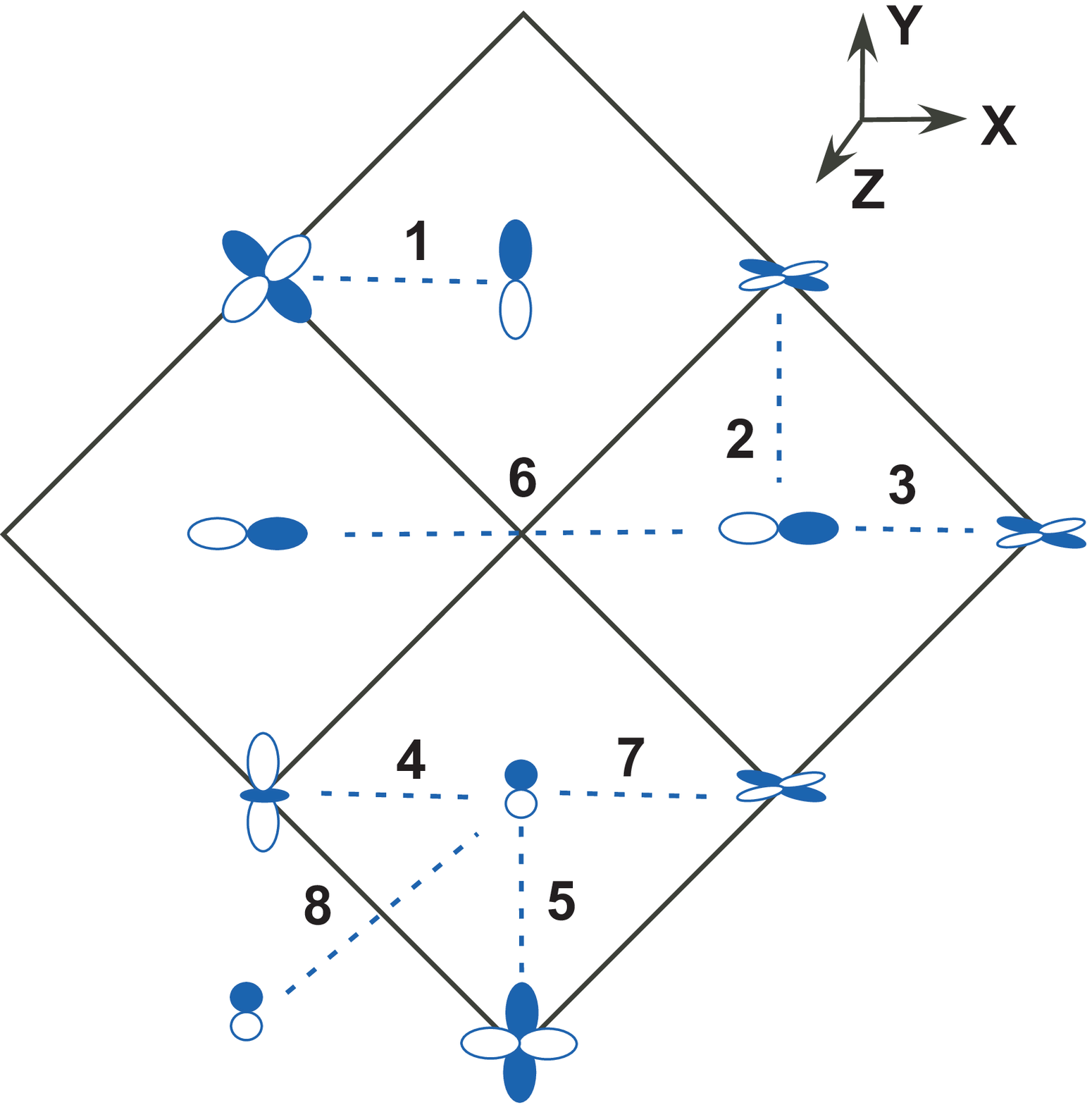}}
  }
  \caption{(Color online)Diagrams for (a) atomic-type MLWFs and (b) strongest hoppings in the system. The diagram is presented on the x-y plane with the most important hoppings labeled. Iron atoms and arsenic atoms are located at the vertices and centers of the
square lattice, respectively. Note that (b) depicts an irreducible subset of hoppings, and the $z$-displacement of As atoms is not shown.}
  \label{fig_hopping_diagram}
\end{figure}


\begin{table}
\caption{Electron hopping $t_{ij}$ and on-site energies $\epsilon_{ij}$ (in eV) matrix
elements calculated from MLWFs. In AFM state,
hopping will be different for iron atoms on different sub-lattices.}
\begin{ruledtabular}
\begin{tabular}{c|c|c|cc}
  &type&PM&\multicolumn{2}{c}{AFM} \\
  \hline
  \multicolumn{5}{c}{on-site} \\
  \hline
  & 3d$_{x^2-y^2}$  & 11.35 & 11.67 & 9.85 \\
  & 3d$_{x(y)z}$    & 11.26 & 11.56 & 9.76 \\
  & 3d$_{xy}$       & 11.18 & 11.49 & 9.60 \\
  & 3d$_{z^2}$      & 11.14 & 11.96 & 10.06 \\
  & 4p$_{x(y)}$     & 9.97  &  9.33 & 9.33 \\
  & 4p$_z$          & 6.35  &  3.17 & 3.17 \\
  \hline
  \multicolumn{5}{c}{hoppings} \\
  \hline
  1 & 3d$_{xy}$-4p$_y$      & 0.79 & 0.80 & 0.62 \\
  2 & 3d$_{xz}$-4p$_x$ $^\prime$ & 0.60 & 0.67 & 0.48 \\
  3 & 3d$_{xz}$-4p$_x$      & 0.81 & 0.79 & 0.74 \\
  4 & 3d$_{z^2-1}$-4p$_z$   & 1.02 & 0.83 & 0.46 \\
  5 & 3d$_{x^2-y^2}$-4p$_z$ & 1.26 & 1.21 & 1.00 \\
  6 & 4p$_x$-4p$_x$         & 0.68 & 0.68 & 0.68 \\
  7 & 3d$_{xz}$-4p$_z$      & 0.49 & $<0.1$ & 0.68 \\
  8 & 4p$_z$-4p$_z$         & 0.17 & 0.55 & 0.55 \\
\end{tabular}
\end{ruledtabular}
\label{tab_hij}
\end{table}
To further understand the physics within the FeAs layers and
connect our calculations to model calculations, we have used the
maximally localized Wannier functions (MLWF) method
\cite{MLWF_1,MLWF_2} to analyze the atomic orbitals which dominate
the electronic structure near the Fermi surface. Sixteen MLWFs,
including 10 d-type MLWFs on Fe and 6 p-type MLWFs on As, have
been used to obtain a tight-binding effective Hamiltonian 
$H_{eff}=\sum_i \epsilon_i c^{\dagger}_ic^{\dagger}_i+\sum_{i,j}t_{ij}c^{\dagger}_ic_j$
by fitting the band structure around $E_F$. These MLWFs are
then used to construct a model Hamiltonian matrix, from which we
can regenerate the band structure using a tight-binding framework
(blue curves in Fig. \ref{fig_band_structure}). In both AFM and PM
cases, the tight-binding band structure fits the DFT band
structure well, showing the validity of our model Hamiltonian. We
show the most important hopping terms in Fig.
\ref{fig_hopping_diagram}, and the corresponding values are listed
in table \ref{tab_hij} together with the on-site energies. 
Due to the S$_4$ symmetry of the FeAs tetrahedra, the Fe
3d orbitals split into 3 non-degenerate (3d$_{x^2-y^2}$, 3d$_{xy}$,
3d$_{z^2}$) and 1 doubly degenerate energy state (3d$_{x(y)z}$) in
both PM and AFM states. Interestingly, the lowest lying 3d$_{z^2}$ state in
PM is the highest in AFM state, leaving the order of other
3 states unchanged. The energy difference between lowest 3d states
and highest 4p orbitals (4p$_{x(y)}$) are $\sim$ 1.2 eV and 1.6 eV in 
PM and AFM state, respectively, which is about the same magnitude as the Cu-O splitting in
cuprates. In the PM state, the strongest hoppings
come from Fe 3d$_{x^2-y^2}$, 3d$_{z^2}$ and As 4p$_z$ orbitals,
followed by the coupling from Fe 3d$_{xz}$, 3d$_{xy}$ and As
4p$_x$ orbitals. Remarkably, the direct hopping between As
4p$_{x(y)}$ in the same x-y plane are as large as the Fe-As hopping. In
the AFM state, since the mirror symmetry within the unit cell is
removed, these hopping matrix elements split into two groups for
spin-up electrons and spin-down electrons on Fe sites,
respectively. Furthermore, the hopping between two neighboring As
4p$_z$ and 4p$_z$ orbitals are greatly enhanced in AFM states. The
direct hoppings between Fe 3d orbitals are finite, but much
smaller compared to Fe-As and As-As hopping in both cases.

\begin{figure}[htp]
  \centering
  \scalebox{0.35}{\rotatebox{270}{\includegraphics{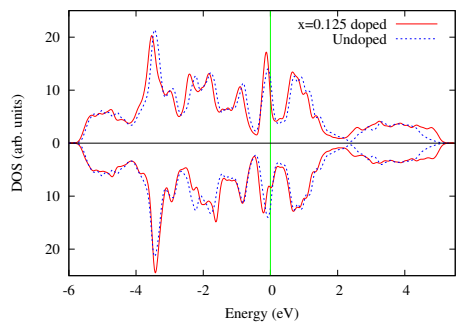}}}
  \caption{(Color online)DOS of \LOFFA\ with $x$=0.125 projected onto FeAs layers.
The $E_F$ is aligned to 0.0 eV. }
  \label{fig_pdos_doped}
\end{figure}
We finally present our calculations on doped \LOFA. The doping is
simulated by substituting an oxygen atom in 8 primitive cells with
a fluorine atom, and then fully relaxing the structure with the optimized lattice
constants of the undoped system. Since each primitive cell contains 2 oxygen atoms, the
doping corresponds to \LOFFA\ with $x=0.0625$. The resulting
material turns out to be still AFM, but with an extra total spin.
With $x$=0.125 doping, we reduced the number of primitive
cells involved by half, so that one oxygen atom in every 4 primitive
cells was replaced with a fluorine atom. The  fluorine substitution has an effect on 
electronic structure which is primarily concentrated in the FeAs layers
at energies near $E_F$, plus an impurity state at $\sim$ 7.5 eV
below $E_F$. We show the comparison of undoped and $x$=0.125 doped system
in Fig. \ref{fig_pdos_doped}. The $x$=0.0625 doped system has a similar
doping effect, but smaller in magnitude. The overall magnetic odering
remains the same, but the magnetic moment is altered by 6\%.

In conclusion, we have performed first-principles calculations for
\LOFA\ and \LOFFA\ systems. An AFM ground state has been found for
undoped \LOFA\ via DFT calculations. We find that the geometry, 
electronic structure and the magnetic state of this system are strongly
related. In both AFM and PM states, the band structures around
Fermi level are derived from Fe 3d and As 4p orbitals, and we have
fitted bands crossing the Fermi surface to tight-binding
Hamiltonians using MLWFs. The parameters for the model
Hamiltonians from the first-principles calculations can be used for
modelling transport, magnetic and superconducting phenomena 
associated with strongly correlated electrons in the system under investigation.
While the system exhibits metallic behavior in DFT calculations,
an inclusion of an on-site energy of 4.5 eV on Fe turns it into a
semiconductor with a gap of 1.0 eV, which implies that the system
is close to a Mott-type insulator. 
Due to the evident proximity of superconductivity to antiferromagnetism
and the Mott transition, we suggest that the system may be a
large-spin analog of the electron-doped cuprates, where AFM and superconductivity coexist.

{\bf Note Added:} Shortly after the first draft of this paper appeared, a linear
spin density wave state was predicted in an electronic structure
calculation \cite{LOFA_SAFM}, and discovered in neutron scattering experiments
\cite{LOFA_Dai_NS}, then extensively studied in \onlinecite{LOFA_Yildrim}.
We have compared the energy of such a magnetic state with the 
sublattice type AFM state discussed in this work, and can confirm
that according to PWSCF calculations it is 109 meV per Fe lower in energy.

\begin{acknowledgments}
This work is supported by DOE DE-FG02-02ER45995, NSF/DMR-0218957 (H.-P. Cheng and C. Cao), and DOE
DE-FG02-05ER46236 (PJH). The authors want to thank D. J. Scalapino for valuable
discussions, and NERSC as well as 
the University of Florida HPC Center for providing computational 
resources and support that have contributed to the research results reported within this paper.
\end{acknowledgments}

\bibliographystyle{apsrev}
\bibliography{PRL}
\end{document}